\PassOptionsToPackage{unicode}{hyperref}
\PassOptionsToPackage{hyphens}{url}
\PassOptionsToPackage{dvipsnames,svgnames,x11names}{xcolor}
\documentclass[
  11pt,
]{article}
\usepackage{xcolor}
\usepackage[margin=1.1in]{geometry}
\usepackage{amsmath,amssymb}
\usepackage{iftex}
\ifPDFTeX
  \usepackage[T1]{fontenc}
  \usepackage[utf8]{inputenc}
  \usepackage{textcomp} % provide euro and other symbols
\else % if luatex or xetex
  \usepackage{unicode-math} % this also loads fontspec
  \defaultfontfeatures{Scale=MatchLowercase}
  \defaultfontfeatures[\rmfamily]{Ligatures=TeX,Scale=1}
\fi
\usepackage{lmodern}
\ifPDFTeX\else
\fi
\IfFileExists{upquote.sty}{\usepackage{upquote}}{}
\IfFileExists{microtype.sty}{% use microtype if available
  \usepackage[]{microtype}
  \UseMicrotypeSet[protrusion]{basicmath} % disable protrusion for tt fonts
}{}
\NewDocumentCommand\citeproctext{}{}
\NewDocumentCommand\citeproc{mm}{%
  \begingroup\def\citeproctext{#2}\cite{#1}\endgroup}
\makeatletter
 \let\@cite@ofmt\@firstofone
 \def\@biblabel#1{}
 \def\@cite#1#2{{#1\if@tempswa , #2\fi}}
\makeatother
\newlength{\cslhangindent}
\newlength{\csllabelwidth}
\newenvironment{CSLReferences}[2] % #1 hanging-indent, #2 entry-spacing
 {\begin{list}{}{%
  \setlength{\itemindent}{0pt}
  \setlength{\leftmargin}{0pt}
  \setlength{\parsep}{0pt}
  \ifodd #1
   \setlength{\leftmargin}{\cslhangindent}
   \setlength{\itemindent}{-1\cslhangindent}
  \fi
  \setlength{\itemsep}{#2\baselineskip}}}
 {\end{list}}
\usepackage{calc}

\usepackage{bookmark}
\IfFileExists{xurl.sty}{\usepackage{xurl}}{} % add URL line breaks if available
\hypersetup{
  pdftitle={Non-Convex Market Makers},
  pdfauthor={Peter Cotton},
  colorlinks=true,
  linkcolor={NavyBlue},
  filecolor={Maroon},
  citecolor={Blue},
  urlcolor={NavyBlue},
  pdfcreator={LaTeX via pandoc}}

\title{Non-Convex Market Makers}
\usepackage{etoolbox}
\makeatletter
\providecommand{\subtitle}[1]{% add subtitle to \maketitle
  \apptocmd{\@title}{\par {\large #1 \par}}{}{}
}
\makeatother
\subtitle{Coherence without Convexity, Friction, and Parallel
Composition}
\author{Peter Cotton}
\date{First version: August 29, 2026. This version: October 31, 2026.}

\begin{document}
\maketitle
\begin{abstract}
Cost-function market makers are always assumed convex, and convexity is
not what makes them coherent. For a path-independent maker the exact
no-arbitrage condition is a chord condition, that every chord slope of
the cost lie in the convex hull of payoffs; convexity enters the
standard axiomatics through a separate requirement, monotone information
incorporation, and dropping it leaves coherent non-convex makers.
Rational flow against such a maker trades the convex envelope, so
non-convexity costs expressiveness rather than soundness, and a
proportional fee of at least the arbitrage depth restores no-arbitrage,
which by conjugation is a bid-ask spread with an exact no-trade interval
for any cost whatever. Combining fee-bearing makers is an infimal
convolution solved by one monotone clearing-price root-find, with sparse
participation and a consolidated limit order book as the aggregate, and
a deep quadratic co-quoter is Moreau smoothing, which preserves
coherence at every depth. Any quoted configuration outside the coherent
hull admits a portfolio priced below its worst payoff, so arbitrageurs
act as decentralized separation oracles, and coherence is relative to
friction and to the cost of finding certificates.
\end{abstract}

\section{Many predictors are already
markets}\label{many-predictors-are-already-markets}

A cost-function market maker posts a potential \(C\) over its inventory
\(q\) of outstanding shares and charges \(C(q+s) - C(q)\) for a fill of
\(s\) shares; prices are gradients of \(C\), and when \(C\) is convex
the standard theory applies
(\citeproc{ref-hanson2003combinatorial}{Hanson 2003};
\citeproc{ref-abernethy2013efficient}{Abernethy et al. 2013}). A run of
published identities makes many learning procedures markets in this
literal sense. Trading against such a maker is
follow-the-regularized-leader with the cost as conjugate regularizer
(\citeproc{ref-chen2010newunderstanding}{Chen and Vaughan 2010};
\citeproc{ref-abernethy2013efficient}{Abernethy et al. 2013}), with
trader wealth as the learning rate
(\citeproc{ref-frongillo2012interpreting}{Frongillo et al. 2012}). A
market of Kelly bettors performs Bayesian model averaging with wealths
as posterior weights (\citeproc{ref-beygelzimer2012kelly}{Beygelzimer et
al. 2012}), and equilibrium prices of utility-maximizing agents
implement mixtures and products of experts
(\citeproc{ref-pennock1997aggregate}{Pennock and Wellman 1997};
\citeproc{ref-storkey2011machine}{Storkey 2011};
\citeproc{ref-storkey2012isoelastic}{Storkey et al. 2012};
\citeproc{ref-barbu2012artificial}{Barbu and Lay 2012}).

Two further identities are used throughout. Precision-weighted
estimation of a common mean is a market of data points. Each observation
is a quadratic maker quoting its value, with capital equal to its
precision, the inverse variance, and the estimate is their merge:
merging makers is the infimal convolution of their costs, under which
liquidity, the inverse of price impact, adds
(\citeproc{ref-bhaskara2023general}{Bhaskara et al. 2023};
\citeproc{ref-barrieu2005inf}{Barrieu and El Karoui 2005}). General
least squares with regressors is not this one-quantity merge:
observation \(i\) contributes the rank-one potential
\((y_i - x_i^\top\beta)^2/(2\sigma_i^2)\) and the potentials add in
parameter space, which asks for vector securities or for eliminating
internal variables from a network of makers, treated in the companion
paper (\citeproc{ref-cotton2026clamped}{Cotton 2026}).

A proximal step is a trade against a fee-bearing maker: the proximal
operator of \(f\lvert\cdot\rvert\) is the soft-threshold, which is the
optimal response to a proportional fee (Lemma 4), and the prox of any
convex \(g\) is the response to a unit quadratic maker charging
\(\tfrac12 s^2 + g(s)\), since
\(\operatorname{prox}_g(x) = \arg\max_s \{x s - \tfrac12 s^2 -
g(s)\}\); the quadratic term is the maker's own curvature, and without
it the trader's problem is a bare conjugate, unbounded for a belief
\(\mu\) with \(\lvert\mu\rvert > f\) in the \(\ell_1\) case. Lemma 4
keeps the base cost \(g_q\) for this reason. The two canonical penalties
are then the two market primitives: ridge is a zero-quoting participant
with capital \(\lambda\), lasso is a fee of \(\lambda\), and the theorem
that regularization is robustness to data perturbation
(\citeproc{ref-elghaoui1997robust}{El Ghaoui and Lebret 1997};
\citeproc{ref-xu2009robustness}{Xu et al. 2009}) acquires a market
reading in which the adversary's budget is priced rather than assumed.

We ask the converse: which predictors are markets, and what does the
market form demand? Serial composition, where markets are chained along
a factorization, is developed in the companion paper
(\citeproc{ref-cotton2026clamped}{Cotton 2026}), and nothing here
depends on it.

The ingredients are classical and the nearest neighbours are close. That
competing makers operate in parallel, that their aggregate is an infimal
convolution, that conjugates add and liquidity adds, and that fees can
be levied on the aggregate, are established in Bhaskara et al.
(\citeproc{ref-bhaskara2023general}{2023}); the risk-sharing form of the
same law is Barrieu and El Karoui (\citeproc{ref-barrieu2005inf}{2005}),
and cost-function markets, their duality and their online-learning
reading are Abernethy et al.
(\citeproc{ref-abernethy2013efficient}{2013}) and Chen and Vaughan
(\citeproc{ref-chen2010newunderstanding}{2010}). The results here are
coherence without convexity and its contact-set consequences; the exact
scalar routing formula with fees inside the costs, its no-trade bands
and sparse participation, and the consolidated-order-book reading;
arbitrage depth, the friction characterization and accessible coherence;
and the synthesis with regularization and incentive closure.

Two payoff regimes appear. In the \emph{bounded} regime a scalar
security settles at \(\varphi(\omega) = \omega \in [-1,1]\), vector
statements substituting the convex hull of payoff vectors; Proposition 1
applies here and bounded worst-case loss is meaningful. In the
\emph{unrestricted} regime the security settles at a real-valued
quantity with payoff hull \(\mathbb{R}\), so the chord condition is
vacuous for finite positions and a quadratic maker
\(C(q) = mq + q^2/(2\lambda)\), whose chord slopes are unbounded, is
admissible; this is the estimation regime of precision-weighted
estimation and of Proposition 10. Coherence statements transfer between
the regimes only through the hull that defines them. Reference
implementations and numerical theorem tests are in the
\texttt{mechanisms} repository.

\section{Coherence is a chord
condition}\label{coherence-is-a-chord-condition}

Throughout, \emph{arbitrage} means a fill whose profit is strictly
positive for every outcome, and \emph{no-arbitrage} the absence of one;
this is the strong or sure-loss convention, and it counts a quote on the
boundary of the payoff hull as coherent even though the fill it permits
is weakly profitable, breaking even on the outcomes that attain the
boundary payoff and gaining on the rest. Under the nonnegative-payoff
convention the statements hold with relative interiors throughout.

\textbf{Proposition 1 (no-arbitrage without convexity).} \emph{For a
path-independent maker (one whose charge depends only on the inventory
endpoints) with cost \(C\), no convexity assumed, there is no
outcome-independent strictly profitable fill from any state if and only
if every chord slope \([C(q+s) - C(q)]/s\) lies in the convex hull of
payoffs: for all \(q, s\),}

\[C(q+s) - C(q) \;\ge\; \min_{\omega}\, \varphi(\omega)\, s
\qquad\text{(scalar: } C \text{ is } 1\text{-Lipschitz).}\]

\textbf{Proof.} A fill \(s\) from state \(q\) has sure profit
\(\min_\omega \varphi(\omega) s - [C(q+s) - C(q)]\); positivity for some
pair \((q,s)\) is exactly the failure of the displayed inequality, and
the inequality for all pairs says every chord slope is supported by the
hull. \(\blacksquare\)

Round trips are refunds for any \(C\), convex or not, because the charge
telescopes over closed paths. When the chord condition fails the exploit
is the accumulation of sure-profit net positions at states where quotes
exit the hull.

Convexity enters the standard axiomatics through \emph{information
incorporation}, the marginal-cost monotonicity condition of Abernethy et
al. (\citeproc{ref-abernethy2013efficient}{2013}), not through
no-arbitrage. The habit of assuming convexity has bundled two different
guarantees: monotone price response to flow, and absence of sure-loss
opportunities. The axiom sets impose the first, convexity follows, and
no-arbitrage is then derived with convexity in hand, so the possibility
of a coherent maker with non-monotone quotes never arises. Proposition 1
separates the guarantees: dropping information incorporation while
keeping the chord condition leaves a coherent non-convex maker,
exhibited numerically in the \texttt{mechanisms} repository. Arbitrage
theory without convexity is developed in a different formalism by
Lépinette and Tran (\citeproc{ref-lepinette2017nonconvex}{2017}).

\section{The market trades the
biconjugate}\label{the-market-trades-the-biconjugate}

Assume for this section that \(C\) is finite and admits an affine
minorant, equivalently that \(C^*\) is proper, and let
\(\hat C = C^{**}\) denote the lower convex envelope
(\citeproc{ref-rockafellar1970convex}{Rockafellar 1970}) and
\(g = C - \hat C \ge 0\) the gap. Chord coherence alone does not give
this. The cost \(C(q) = -\alpha\lvert q\rvert\) with \(\alpha < 1\) is
\(\alpha\)-Lipschitz and so chord-coherent for the hull \([-1,1]\), yet
an affine minorant would need slope at most \(-\alpha\) and at least
\(\alpha\) at once; hence \(C^* \equiv +\infty\),
\(C^{**} \equiv -\infty\), the gap is undefined, and at \(q = 0\),
\(\mu = 0\) the trader's supremum \(\sup_s \alpha\lvert s\rvert\) is
already infinite. Coherence bounds what a single fill can extract for
sure; it does not bound speculative profit against a belief.

\textbf{Proposition 2 (contact and pass-through).} \emph{Let \(C\) be as
above and let a myopic risk-neutral trader have believed mean \(\mu\) in
the hull with \(C^*(\mu) < \infty\), facing the maker at state \(q\).
Then the trader: (i) has every attained optimal fill landing on the
contact set \(\{C = \hat C\}\), an optimum existing whenever
\(x \mapsto C(x) - \mu x\) attains its infimum (for instance when it is
coercive); and (ii) earns maximal expected profit}

\[\Pi_C(q,\mu) \;=\; \Pi_{\hat C}(q,\mu) + g(q)\]

\emph{as an identity of extended-real suprema, the envelope profit plus
the gap at the starting state.}

\textbf{Proof.}
\(\sup_s \mu s - [C(q+s) - C(q)] = \sup_x [\mu x - C(x)] - \mu q
+ C(q)\), so the profit is \(C^*(\mu)\) up to terms in \(q\) alone.
Since conjugation is invariant under biconjugation,
\(\hat C^{\,*} = C^{***} = C^*\), the two suprema agree whether or not
either is attained, which gives (ii) after adding and subtracting
\(\hat C(q)\). For (i), a maximizer of an affine function minus \(C\) is
a point where an affine minorant touches \(C\), hence touches
\(\hat C\), so any attained optimum lies in the contact set.
\(\blacksquare\)

Attainment can fail. The chord-coherent cost \(C(x) = \arctan x\) has
lower convex envelope the constant \(-\pi/2\) and hence empty contact
set; at \(\mu = 0\) the supremum \(\pi/2\) is approached only as
\(x \to -\infty\). When the contact set is empty, statements about flow
landing on contact points and off-contact states being transient
describe a limit that no fill realizes.

The maker is value-equivalent to its convex envelope for an
unconstrained one-shot risk-neutral optimizer, and no more than that:
the optimal endpoints of \(C\) are exactly the envelope-optimal
endpoints lying in the contact set, so \(C\) deletes the envelope's
off-contact optima, and partial trades, tie-breaking, capital
constraints and noisy flow distinguish the two immediately. Concave
stretches are unquotable intermediate \emph{inventory} states that
rational flow jumps across. Whoever lands inside one (noise) overpays
the gap at the landing state, and by (ii) the next rational trader
recoups it, so the maker is a conduit keeping envelope differences over
any rational-to-rational span, and off-contact states are transient. The
trade-set analogue for constant-function market makers (CFMMs, the
decentralized-exchange design) is the canonical concave trading
function: an arbitrary invariant is behaviorally equivalent to a concave
one (\citeproc{ref-angeris2024geometry}{Angeris et al. 2024}), with the
limits of concavification mapped by Frongillo et al.
(\citeproc{ref-frongillo2024axiomatic}{2024}).

The gap is the opposite of a gap in prices. Between contact points
\(q_1 < q_2\) the envelope is affine with a single supporting slope
\(\mu_0\), so the marginal price is constant across the excluded
inventory range: as belief crosses \(\mu_0\) the optimal inventory jumps
from \(q_1\) to \(q_2\) while the price does not move. In the
convexified supply correspondence this is an inventory jump of size
\(q_2 - q_1\) at the supporting price \(\mu_0\), a depth spike rather
than a missing price range. Against the original non-convex cost the
jump is generally indivisible: for the symmetric double well of
Proposition 9 the full trade from \(-c\) to \(c\) costs nothing while
the half trade from \(-c\) to \(0\) costs \(\alpha c\), so the prefixes
of the block are not executable at \(\mu_0\) and this is a lumpy
endpoint trade, not a conventional divisible level. Non-convexity of
\(C\) also does not by itself make an implied density bimodal: \(C^*\)
is convex, so any factor read as \(\exp(-C^*)\) is log-concave. The gap
is neither a hole in the book nor a multimodal quote, and the bimodal
implied densities documented around binary events
(\citeproc{ref-melick1997crude}{Melick and Thomas 1997};
\citeproc{ref-clark2017brexit}{Clark and Amen 2017}) are a different
object.

\section{Frictions price the remaining
failure}\label{frictions-price-the-remaining-failure}

After Proposition 1 the only failure mode is chord slopes exiting the
hull.

\textbf{Proposition 3 (fees buy bounded incoherence).} \emph{If chord
slopes exit the hull by at most \(\varepsilon\) per unit,
i.e.~\(C(q+s) - C(q) \ge
\min_\omega \varphi(\omega)s - \varepsilon\lvert s\rvert\), then a
proportional fee \(f \ge \varepsilon\) restores no-arbitrage.}

\textbf{Proof.} Sure profit with the fee is
\(\min_\omega \varphi(\omega)s - \Delta C - f\lvert s\rvert \le
(\varepsilon - f)\lvert s\rvert \le 0\). \(\blacksquare\)

This is the mechanism-level counterpart of a theorem of mathematical
finance: under regularity conditions such as sticky paths or full
support, arbitrarily small proportional transaction costs restore
no-arbitrage for a large class of processes that are arbitrageable
frictionlessly, with a consistent price system inside the spread as
certificate (\citeproc{ref-guasoni2006transaction}{Guasoni 2006};
\citeproc{ref-guasoni2010ftap}{Guasoni et al. 2010}). In prediction
markets the precedent is fees sized to expected arbitrage profit
restoring bounded loss under privacy noise
(\citeproc{ref-cummings2016privacy}{Cummings et al. 2016};
\citeproc{ref-frongillo2018private}{Frongillo and Waggoner 2018}). Call
the smallest \(\varepsilon\) for which the hypothesis of Proposition 3
holds the cost's \emph{arbitrage depth}: the worst per-unit excursion of
its chord slopes beyond the payoff hull, and so the minimum viable
half-spread (the quoted spread being \(2f\), since bid and ask are
\(m \mp f\)). A venue quoting a wide spread to cover a deeply incoherent
cost is arbitrage-free but uninformative in proportion: the market
prices the model's incoherence as uncertainty.

\section{A linear fee is a bid-ask
spread}\label{a-linear-fee-is-a-bid-ask-spread}

Why a fee at all? A cost-function maker's charge telescopes, so round
trips are free: the maker earns nothing on uninformed flow and cannot
recover from volume its adverse-selection losses, the losses to
better-informed traders. That no state-dependent cost can charge a round
trip, and that a path-dependent volume charge repairs it, is due to
Othman and Sandholm (\citeproc{ref-othman2012profitcharging}{2012}); the
linear charge \(f\lvert s\rvert\) is their volume levy in its simplest
convex form, chosen because it conjugates in closed form. Write
\(g_q(s) = C(q+s) - C(q)\) for a convex \(C\) and \(m = C'(q)\) for the
marginal price.

\textbf{Lemma 4 (fee--spread duality).} \emph{Let
\(T_f(x) = \operatorname{sign}(x)\max(\lvert x\rvert - f,\, 0)\) denote
the soft-threshold. Then the fee-bearing cost \(\tilde C_q = g_q +
f\lvert\cdot\rvert\) has conjugate}

\[\tilde C_q^*(p) \;=\; g_q^*\!\big( m + T_{f}(p - m) \big),\]

\emph{zero if and only if \(\lvert p - m\rvert \le f\).}

\textbf{Proof.} The conjugate of a sum of closed proper convex functions
with overlapping relative interiors of domains is the infimal
convolution of the conjugates
(\citeproc{ref-rockafellar1970convex}{Rockafellar 1970}, Thm. 16.4), and
the conjugate of \(f\lvert\cdot\rvert\) is the indicator of \([-f, f]\).
Hence \(\tilde C_q^*(p) = \min_{\lvert u\rvert \le f} g_q^*(p - u)\),
the minimum of the convex function \(g_q^*\) over \([p - f,\, p + f]\).
Since \(g_q^* \ge 0\) with equality exactly at \(m\), the minimum is
attained at the projection of \(m\) onto the interval, which is
\(m + T_{f}(p - m)\). \(\blacksquare\)

The maker quotes \(\mathrm{ask} = m + f\) and \(\mathrm{bid} = m - f\)
and trades nothing in between: a proportional fee is a bid-ask spread,
the same duality by which a proportional transaction cost confines the
pricing functional to the bid-ask band
(\citeproc{ref-jouini1995transaction}{Jouini and Kallal 1995}). Without
convexity the exact description needs no conjugation at all:

\textbf{Lemma 5 (the no-trade interval, exactly).} \emph{For any cost
\(C\) (no convexity assumed), state \(q\), and fee \(f \ge 0\), write
\(d_q(s) = [C(q+s) - C(q)]/s\) for the chord slope. A belief \(\mu\)
admits no profitable trade if and only if}

\[\sup_{s<0} d_q(s) - f \;\le\; \mu \;\le\; \inf_{s>0} d_q(s) + f,\]

\emph{so the set of no-trade beliefs is exactly this interval
intersected with the payoff hull. For differentiable convex \(C\) both
bounds equal \(C'(q)\) and the interval is Lemma 4's band
\([m - f, m + f]\); at a convex kink the bounds are the one-sided
derivatives, so the interval is
\([\partial^- C(q) - f,\ \partial^+ C(q) + f]\) and the kink contributes
a spread of its own even at \(f = 0\). Write
\(\Delta_q = \sup_{s<0} d_q - \inf_{s>0} d_q\). The untruncated interval
is non-empty if and only if \(2f \ge \Delta_q\); if in addition \(C\) is
chord-coherent, both bounds lie in the hull, so their midpoint does and
the intersection with the hull is non-empty under the same condition. If
\(C\) is chord-coherent and Proposition 2's supremum is attained, then
off the contact set \(\Delta_q > 0\).}

\textbf{Proof.} No profitable trade means \(\mu s \le C(q+s) - C(q) +
f\lvert s\rvert\) for all \(s\). Dividing by \(s > 0\) gives
\(\mu \le d_q(s) +
f\); dividing by \(s < 0\) reverses the inequality to
\(\mu \ge d_q(s) - f\); together these are the displayed interval, whose
non-emptiness is \(\sup_{s<0} d_q - f \le \inf_{s>0} d_q + f\). Under
chord coherence both one-sided bounds are limits of chord slopes and so
lie in the hull, which is convex, so the midpoint of a non-empty
interval lies in the hull. For the last claim, off the contact set
Proposition 2 gives every belief a strictly positive frictionless
profit, so the \(f = 0\) interval is empty. \(\blacksquare\)

Without the hull intersection the statement fails. Without chord
coherence \(C(q) = 100q\) has \(\Delta_q = 0\), so the untruncated
interval \(\{100\}\) is non-empty for every \(f \ge 0\), yet no
admissible belief in \([-1,1]\) declines to trade: the maker is
arbitraged from every state.

So friction does not merely widen an existing spread: at
\(2f \ge \Delta_q\) a state in the excluded inventory range becomes
tenable, admitting no profitable trade at some admissible belief.
Non-convexity makes an inventory range transient, frictionless rational
flow jumps it, and a large enough spread lets the maker rest inside it.
For \(C(q) = a\sin q\) at \(q = 0\), a state \(a\) above its flat
envelope, the belief \(\mu = 0\) admits no profitable trade exactly when
\(f \ge a\), though every belief profits there frictionlessly.

\section{Makers in parallel}\label{makers-in-parallel}

Let makers \(i = 1..n\) hold inventories \(q_i\) with costs \(C_i\) that
are differentiable and strictly convex, with \(C_i'\) onto the price
range of interest so that \((C_i')^{-1}\) is defined there; fees \(f_i\)
of their choosing; and write \(m_i = C_i'(q_i)\) for maker \(i\)'s
marginal price and \(\mathrm{ask}_i = m_i + f_i\),
\(\mathrm{bid}_i = m_i - f_i\) for its quotes, as in Lemma 4. Without
differentiability and strict convexity the supply curves are set-valued
and the statements hold with \((C_i')^{-1}\) read as a subgradient
correspondence.

One normalization is needed before the merged object is a maker. The
infimal convolution of the incremental costs need not vanish at zero:
for two zero-fee quadratic makers \(g_i(s) = m_i s + s^2/(2\lambda_i)\),

\[(g_1 \,\square\, g_2)(0) \;=\; -\frac{(m_1-m_2)^2}
{2(1/\lambda_1 + 1/\lambda_2)} \;<\; 0\]

whenever the makers' quotes differ, because zero net external demand
still admits a profitable internal cross-trade. We therefore assume the
component inventories have been cleared against each other and the
effective cost normalized to vanish at the resulting origin; otherwise
evaluating at \(\Delta = 0\) repeatedly appears to pay the same internal
arbitrage again and again.

\textbf{Lemma 6 (combination).} \emph{Let
\(\tilde C = \tilde C_1 \,\square\, \cdots \,\square\, \tilde C_n\) and
define each maker's supply}

\[s_i(p) \;=\;
\begin{cases}
(C_i')^{-1}(p - f_i) - q_i, & p \ge \mathrm{ask}_i,\\[2pt]
0, & \mathrm{bid}_i < p < \mathrm{ask}_i,\\[2pt]
(C_i')^{-1}(p + f_i) - q_i, & p \le \mathrm{bid}_i,
\end{cases}\]

\emph{each non-decreasing in \(p\). Fix a demand \(\Delta\) for which a
clearing price \(p^*\) with \(\sum_i s_i(p^*) = \Delta\) exists. Then:
(i) \(\tilde C^* = \sum_i \tilde C_i^*\), a sum of soft-thresholded
profit functions; (ii) the split \(s_i = s_i(p^*)\) attains
\(\tilde C(\Delta)\), and any optimal split satisfies
\(C_i'(q_i + s_i) + f_i \operatorname{sign}(s_i) = p^*\) for
\(s_i \ne 0\) and \(\lvert p^* - m_i \rvert \le f_i\) for \(s_i = 0\);
(iii) the split is sparse: every maker whose quote band strictly
contains \(p^*\) trades exactly zero.}

\textbf{Proof.} (i) is the conjugate-sum identity applied to the
convolution (\citeproc{ref-rockafellar1970convex}{Rockafellar 1970}).
For (ii), the split is feasible by choice of \(p^*\), and
\(p^* \in \partial \tilde C_i(s_i(p^*))\) for every \(i\): when
\(s_i(p^*) \ne 0\) the subgradient is
\(C_i'(q_i+s_i) + f_i\operatorname{sign}(s_i) = p^*\), and when
\(s_i(p^*) = 0\) it is the interval
\([m_i - f_i,\, m_i + f_i] \ni p^*\). A common multiplier certifying
every coordinate is exactly the optimality condition for
\(\min\{\sum_i \tilde C_i(s_i) : \sum_i s_i = \Delta\}\). (iii) restates
the zero branch. \(\blacksquare\)

The computation is a scalar monotone root-find whatever \(n\) is, and
the fee costs nothing beyond a horizontal shift of each supply curve.
The \(\lvert s\rvert\) terms are an \(\ell_1\) penalty, so sparsity
arrives for the same reason it does in the lasso, and for the same
reason proportional transaction costs produce no-trade regions and
sparse portfolios (\citeproc{ref-olivaresnadal2018robust}{Olivares-Nadal
and DeMiguel 2018}). A small trade routes to whichever maker posts the
best quote, and to that maker alone unless the best quote is tied; a
growing trade pushes that maker's fee-adjusted marginal price through
the next band and spills over, consuming makers in quote-price order,
which coincides with fee order only when their marginal prices agree.

\textbf{Remark 7 (zero fees).} \emph{With \(f_i \equiv 0\) the
convolution reduces to the fee-free merge: conjugate regularisers add,
and for a perspective family \(C_b(q) = b\,C_1(q/b)\) liquidity adds,
\(C_{b_1} \square C_{b_2} = C_{b_1+b_2}\). This is the aggregation law
of Bhaskara et al. (\citeproc{ref-bhaskara2023general}{2023}), in
risk-measure form Barrieu and El Karoui
(\citeproc{ref-barrieu2005inf}{2005}). For quadratic makers the merge is
Gaussian fusion: the merged quote is the precision-weighted mean of the
makers' quotes and the merged liquidity the sum of their precisions.}

\textbf{Corollary 8 (the order book).} \emph{The aggregate supply
\(S(p) = \sum_i s_i(p)\) is non-decreasing, identically zero on
\(\big(\max_i \mathrm{bid}_i,\ \min_i \mathrm{ask}_i\big)\) when that
interval is non-empty, flat on any open interval contained in the
interiors of all the bands, and continuous and strictly increasing
wherever some maker is active. If the active makers are in addition
\(C^2\) with \(C_i'' > 0\) and \(p\) is away from band boundaries, the
local depth is
\(S'(p) = \sum_{i\ \mathrm{active}} 1/C_i''(q_i + s_i(p))\).}
Differentiability needs that extra hypothesis: the maker \(C(q) = q^4\)
at \(q_i = -1\) is differentiable and strictly convex with supply
\(S(p) = (p/4)^{1/3} + 1\), active at \(p = 0\) and not differentiable
there. Read as a market: best bid and ask are the tightest quotes,
\(S(p)\) is the cumulative quantity executable up to price \(p\) with
local depth \(S'(p)\) where it exists, and large orders walk the levels.
The aggregate of linear-fee makers is a consolidated limit order book,
and in producer-theory terms Lemma 6 is Marshall's horizontal summation
of firm supply curves (\citeproc{ref-marshall1890principles}{Marshall
1890}; \citeproc{ref-mascolell1995microeconomic}{Mas-Colell et al.
1995}) with the reversibility of share production patched by the fee.
The economics of the book assembled from competing liquidity suppliers
is Glosten (\citeproc{ref-glosten1994limit}{1994}), with convergence of
strategic schedules in Biais et al.
(\citeproc{ref-biais2000competing}{2000}).

\textbf{Proposition 9 (a deep co-quoter is Moreau smoothing).}
\emph{Merging a maker with cost \(C\) with a quadratic co-quoter of cost
\(s^2/(2\lambda)\) yields the venue with cost the Moreau envelope
\(e_\lambda C(x) = \min_y C(y) + (x-y)^2/(2\lambda)\). Then: (i)
coherence is preserved: if the chord slopes of \(C\) lie in \([a, b]\),
so do those of \(e_\lambda C\), for every \(\lambda\); (ii) at any point
where two distinct minimizing branches coexist (a crossing), the
downward jump in marginal price is exactly the separation of the
competing minimizers divided by \(\lambda\), at most \(D/\lambda\) when
all competing minimizers lie in an interval of diameter \(D\); (iii) for
the symmetric double well
\(C(y) = \alpha \min(\lvert y - c\rvert, \lvert y + c\rvert)\) with
\(\alpha < 1\), the jump at \(x = 0\) equals \(2\alpha\) while
\(\lambda\alpha < c\) and \(2c/\lambda\) thereafter, so it is constant
in the depth until the co-quoter is deep enough to reach across the
wells. The bound in (ii) measures crossing defects only: the envelope
can be smoothly non-convex with a unique minimizer everywhere, as for
\(C(y) = a \sin y\) with \(\lambda a < 1\), where
\((e_\lambda C)''(x) = C''(y^*)/(1 + \lambda C''(y^*))\) is negative
wherever \(C''\) is.}

\textbf{Proof.} The merge is infimal convolution, and inf-convolution
with the quadratic is the Moreau envelope
(\citeproc{ref-rockafellar1970convex}{Rockafellar 1970}). For (i), no
differentiability is needed: for \(h > 0\), substituting \(y = z + h\),

\[e_\lambda C(x+h) = \inf_z \Big\{ C(z+h) + \frac{(x-z)^2}{2\lambda}
\Big\} \;\le\; e_\lambda C(x) + b\,h,\]

since \(C(z+h) - C(z) \le b h\) for every \(z\); the reverse
substitution gives the lower bound \(a h\), so every chord slope of
\(e_\lambda C\) lies in \([a, b]\). For (ii), at a crossing \(x_0\) with
competing minimizers \(y_1^* < y_2^*\) the branch slopes are
\((x_0 - y_i^*)/\lambda\), so the drop is exactly
\((y_2^* - y_1^*)/\lambda\). For the smooth case, differentiate the
first-order condition \(y^* = x - \lambda C'(y^*)\) to get
\(dy^*/dx = 1/(1 + \lambda C''(y^*))\) and hence the displayed
curvature. For (iii), the well's competing minimizers are at
\(\pm\lambda\alpha\) while \(\lambda\alpha < c\), giving separation
\(2\lambda\alpha\) and jump \(2\alpha\), and at \(\pm c\) once the
co-quoter reaches the wells, giving \(2c/\lambda\). \(\blacksquare\)

Part (i) is the durable statement: adding depth reshapes a non-convex
maker without ever creating arbitrage, at any \(\lambda\). Depth does
not, however, fill the excluded range at finite depth. For the double
well \(e_\lambda C(\pm c) = 0\) while
\(e_\lambda C(0) = \alpha c - \lambda\alpha^2/2\) for
\(\lambda\alpha < c\) and \(c^2/(2\lambda)\) after, strictly positive
for every finite \(\lambda\); since \(e_\lambda C \ge 0\) vanishes at
\(\pm c\), its convex envelope is zero at the midpoint, so the midpoint
stays strictly off contact at every depth and the gap merely decays like
\(1/\lambda\).

Depth attenuates the envelope gap and regularizes crossing geometry
without creating chord incoherence; it does not close the off-contact
interval, restore convexity, or make every state reachable. Crossing
kinks close at rate \(1/\lambda\) only where the competing minimizers
stay within a \(\lambda\)-independent diameter; the double well shows
what happens otherwise, its jump holding at \(2\alpha\) until the depth
exceeds \(c/\alpha\). Smooth concave stretches lie outside the bound's
scope entirely. Here \(\lambda\) is the co-quoter's liquidity: its price
impact is \(1/\lambda\), so large \(\lambda\) means a deep book, not a
strong pull.

The two repairs are the two market primitives again: friction
(\(\ell_1\), the fee) and participant depth (\(\ell_2\), capital), lasso
and ridge. The identification maps the penalty terms, not the
statistical procedures; it is a dictionary of primitives, not an
equivalence of estimators.

\section{Self-set fees and
adaptivity}\label{self-set-fees-and-adaptivity}

Nothing requires the fees to be administered. Each maker may quote its
own \(f_i\), and routing then disciplines the posted menu: a maker
quoting inside the prevailing spread becomes the best quote and takes
the flow first, a maker quoting behind the best price receives nothing
until the book in front of it is consumed, and a maker quoting too tight
is picked off by informed flow.

The discipline is a property of Lemma 6's routing, not of its
minimization. The infimal convolution minimizes over allocations \(s_i\)
with the fees held fixed; it does not minimize over the fees, and
nothing in it constitutes a game in which makers choose \(f_i\). What
the surviving fee would be, and whether it is the competitive
adverse-selection charge of classical microstructure
(\citeproc{ref-glosten1985bidask}{Glosten and Milgrom 1985};
\citeproc{ref-biais2000competing}{Biais et al. 2000}), requires the
strategic quote game, which is not solved here.

Optimization already pays these frictions: a proximal step charges
\(\lVert\Delta\theta\rVert^2/(2\eta)\) per move, a trust region is an
infinite fee outside a band, weight decay is a zero-quoting participant.
In market language the stabilizers of non-convex training are the
market's two repairs: friction prices chord excursions (Proposition 3)
and depth attenuates the envelope gap (Proposition 9); non-convexity by
itself is not exploitability (Proposition 1). Path-dependent optimizers
correspond to makers whose quotes depend on flow history, the
adaptive-liquidity class, where path independence is given up: no maker
combines path independence, translation invariance (complete-bundle
prices summing to the sure payoff) and liquidity sensitivity
(\citeproc{ref-othman2013practical}{Othman et al. 2013}), the adaptive
class is axiomatized by Li and Wortman Vaughan
(\citeproc{ref-li2013adaptive}{2013}) with the homogeneous-risk-measure
characterization in Othman and Sandholm
(\citeproc{ref-othman2011liquidity}{2011}), no trade-history maker
achieves every desideratum at once
(\citeproc{ref-abernethy2014vpm}{Abernethy et al. 2014}), and liquidity
selection itself can be run as online learning
(\citeproc{ref-nueve2026adaptiveliquidity}{Nueve et al. 2026};
\citeproc{ref-nueve2025smooth}{Nueve and Waggoner 2025}).

\section{Inconsistency is arbitrage}\label{inconsistency-is-arbitrage}

The results so far concern one venue and its aggregate. The same convex
geometry says when \emph{several} venues, quoting related securities,
are jointly incoherent, and it turns each incoherence into a portfolio
that pays whoever finds it.

\textbf{Proposition 10 (quote inconsistency is a sure profit).}
\emph{Let \(x\) be a vector of unrestricted real random variables and
let separate venues quote securities paying the products \(x_i x_j\) for
every pair, the diagonal quoted at one, at fixed linear prices
executable in the size traded; assemble the quotes into a matrix \(P\)
with unit diagonal. The quotes admit a joint distribution iff
\(P \succeq 0\) (a Gaussian with covariance \(P\) witnesses
sufficiency). Otherwise let \(w\) be a unit eigenvector of a negative
eigenvalue \(\lambda_{\min}\) and buy \(w_i^2\) units of security \(ii\)
and \(2 w_i w_j\) units of each security \(ij\), \(i < j\). The bundle
costs \(w^\top P w = \lambda_{\min} < 0\) and pays
\((w^\top x)^2 \ge 0\): a sure profit of \(\lvert\lambda_{\min}\rvert\).
On three coordinates the quoted pairs are exactly the edges of a
triangle, so the cycle reading is literal.}

This is the second-moment coherence condition, prices of products must
form a PSD matrix (\citeproc{ref-daspremont2005market}{{d'Aspremont}
2005}), read as cycle consistency. Arbitrage-enforced coherence is an
old idea. Coherence is no-arbitrage (\citeproc{ref-nau1991arbitrage}{Nau
and McCardle 1991}); Pennock and Wellman
(\citeproc{ref-pennock1996marketbayes}{1996}) built arbitrageur agents
enforcing the additivity identities of a Bayes-net economy, with
equilibrium prices equal to the network's probabilities and distributed
bidding as distributed inference; the combinatorial literature removes
the arbitrage algorithmically over the marginal polytope
(\citeproc{ref-kroer2016arbitrage}{Kroer et al. 2016}); and Saguillo et
al. (\citeproc{ref-saguillo2025arbitrage}{2025}) measure Polymarket
arbitrageurs extracting \$39.6M enforcing logical coherence across
markets.

Two boundaries of the proposition mark the general picture. If only a
subgraph of pairs is quoted, the realizability condition is not positive
semidefiniteness of a filled-in matrix but PSD-completability of the
partial one (\citeproc{ref-grone1984completions}{Grone et al. 1984}),
and the arbitrage bundle must be supported on quoted pairs. If the
coordinates are constrained, say binary \(x_i \in \{\pm 1\}\), the
feasible quote set is the cut polytope, a strict subset of the PSD body
(\citeproc{ref-deza1997cuts}{Deza and Laurent 1997}), so quotes can be
positive semidefinite yet jointly unrealizable. In every case the
coherent set is the convex hull of attainable payoffs, and Proposition
11 is the universal form.

The scope of Proposition 10 is in any case one class of consistency
failure: locally quoted beliefs that cannot be embedded in any joint
distribution, detected here at second moments. Proposition 10 is a
second-moment arbitrage result, not a theorem about loopy belief
propagation in general: not every loopy-propagation error takes this
form, and the reading of arbitrageurs as the loop correction is a
conjecture about dynamics, not modelled here. The class it covers is, in
a market, free money. Whether flow that harvests it converges, and to
the true marginal, to a surrogate of the kind loopy propagation
converges to, or to something the liquidity profile selects, is open.

The PSD cone is not special. Coherent price sets are convex (they are
convex hulls of payoff vectors, or their conic images), and separation
turns every exterior point into a trade:

\textbf{Proposition 11 (arbitrage is separation).} \emph{Let \(K\) be
the closed convex hull of the attainable payoff vectors (or its image
under a linear security map), so that
\(\inf_{z \in K} \langle y, z\rangle \le \langle y,
\varphi(\omega)\rangle\) for every outcome \(\omega\) and portfolio
\(y\), and let the venue quote fixed linear prices \(x\), executable in
the size traded. If \(x \notin K\), any separating functional \(y\) with
\(\langle y, x\rangle < \inf_{z \in K} \langle y, z\rangle\) is a
portfolio whose price is strictly less than its realized payoff in every
state: a sure profit. Proposition 10 is the instance
\(K = \{P \succeq 0\}\).}

\emph{For a nonlinear cost-function maker the conclusion is local. There
\(x\) is the current gradient and a trade of size \(\delta y\) costs
\(\delta \langle y, x \rangle + o(\delta)\), since the marginal price
moves as the trade is filled, so a separating margin exceeding
\(f\lVert y\rVert\) yields sure profit for all sufficiently small
\(\delta\), at a rate rather than in the size of \(y\); a margin below
that is suppressed by Proposition 3, and market impact then sets how
large a position the certificate supports.}

The proof is the separating-hyperplane theorem read as a trade
(\citeproc{ref-nau1991arbitrage}{Nau and McCardle 1991};
\citeproc{ref-daspremont2005market}{{d'Aspremont} 2005}); the
payoff-hull hypothesis upgrades the separation certificate to an
arbitrage, and for an abstract consistency set the construction yields
the certificate only. In an ordinary numerical method an infeasibility
is a residual to be driven down by the algorithm; in a market it is a
payoff, and whoever finds it is paid to act as the separation oracle.
Arbitrageurs are decentralized separation oracles, and the friction of
Proposition 3 sets the tolerance below which infeasibility is allowed to
persist.

\section{Closing}\label{closing}

The characterization is for the cost-based class. Call a predictor
\emph{cost-based} if it is specified by a path-independent potential
\(C\) over a security inventory, and call friction of size \(f\)
\emph{permitted} if the mechanism may charge \(f\lVert s\rVert\) on a
fill \(s\). For vector inventories arbitrage depth is measured against
the same norm,

\[\varepsilon \;=\; \sup_{q,\,s \ne 0} \frac{\big[\inf_{z \in K}
\langle z, s\rangle - C(q+s) + C(q)\big]_+}{\lVert s\rVert},\]

the scalar case being \(\lVert\cdot\rVert = \lvert\cdot\rvert\). A
cost-based predictor admits a coherent market implementation if and only
if its arbitrage depth is dominated by the permitted friction. The
equivalence is close to definitional, since arbitrage depth is the
maximal normalized violation, and it fixes vocabulary. The content is
that the depth is finite for costs no axiom set in the literature
admits, and that Propositions 1 to 3 say what finiteness costs.

Chord coherence is the \(f = 0\) special case, and convexity is the
separate, further property of monotone information incorporation. Three
words are kept apart throughout: \emph{coherence} (chord slopes stay in
the payoff hull), \emph{convexity} (marginal prices respond
monotonically to flow), and \emph{expressiveness} (which inventory
states rational flow can leave the maker in). Within the proper-envelope
class, and for beliefs whose response is well posed, non-convexity
restricts the rationally reachable inventory states to the contact set:
\(C\) and \(\hat C\) give the same optimal value, with

\[\operatorname*{argmax}_x\{\mu x - C(x)\} \;=\;
\operatorname*{argmax}_x\{\mu x - \hat C(x)\} \cap \{C = \hat C\},\]

so the original cost deletes the envelope's off-contact optimizers, and
from an off-contact state the next rational trader extracts the gap
\(g(q)\). Coherence alone promises no more than this, since
\(C(q) = -\alpha\lvert q\rvert\) is coherent with unbounded speculative
profit.

The target operation is \emph{marketization}; its single-maker and
parallel parts are Propositions 1 to 11, and its serial part is the
subject of the companion paper. Given an operator \(T\) from inputs to
outputs, a marketization is a mechanism whose clearing computes \(T\);
whose local contributions compose, in parallel and in series, to compute
composite operators; in which inconsistent contributions create
exploitable trades (Proposition 11); in which friction bounds how much
inconsistency can persist (Proposition 3); and in which every
participant pays to perturb the computation. One estimator becomes one
maker and combining evidence becomes parallel composition; composing
conditional operators is serial composition, developed separately. A
statistical procedure specifies how information would be combined if
supplied honestly; its marketization implements the same operator
against self-interested sources. In this sense a market is the incentive
closure of a predictor: the computation, plus the dual certificates of
its constraints, plus payment to whoever holds one. Duality alone
supplies the middle column of

\[\begin{array}{lll}
\text{computation} & \text{certificate} & \text{payment}\\
\text{primal solve} & \text{dual multiplier} & \text{none}\\
\text{verified computation} & \text{proof of violation} & \text{none or fixed}\\
\text{incentive closure} & \text{separating portfolio} & \text{the certificate's value}
\end{array}\]

and the closure adds the third: a marketized computation is
self-policing in proportion to the value of its own errors, with
friction the exchange rate between tolerated error and the cost of
participation.

The definition has neighbors in four literatures, and each holds one
column. Markets that compute are old: a combinatorial auction cleared
airport slots by integer programming
(\citeproc{ref-rassenti1982slots}{Rassenti et al. 1982}), electricity
markets pay the dual variables of the dispatch program as nodal prices
(\citeproc{ref-schweppe1988spot}{Schweppe et al. 1988}), and
market-oriented programming solved distributed allocation by computing
competitive equilibria of artificial economies
(\citeproc{ref-wellman1993market}{Wellman 1993}).

Algorithmic mechanism design asks when optimization is compatible with
incentives (\citeproc{ref-nisan2001algorithmic}{Nisan and Ronen 2001}),
and the classification of what can be computed incentive-compatibly was
posed as a program by Feigenbaum and Shenker
(\citeproc{ref-feigenbaum2002distributed}{2002}). The middle column
carries a lower bound: any communication protocol realizing an efficient
allocation must reveal supporting prices
(\citeproc{ref-nisan2006communication}{Nisan and Segal 2006}), so
certificates are not optional. The third column exists in practice
without the duality: fraud-proof systems pay whoever exhibits an invalid
state transition (\citeproc{ref-teutsch2019truebit}{Teutsch and
Reitwießner 2019}; \citeproc{ref-kalodner2018arbitrum}{Kalodner et al.
2018}), for discrete transitions rather than convex programs. And the
friction bound has an incentive-free ancestor: the auction algorithm's
\(\varepsilon\)-complementary slackness permits suboptimality at most
\(n\varepsilon\), vanishing for \(\varepsilon < 1/n\) on integer data
(\citeproc{ref-bertsekas1992auction}{Bertsekas 1992}); Proposition 3 is
the same shape of bound with the \(\varepsilon\) charged to traders. The
connective tissue, the closure operation itself, serial composition of
mechanisms, and the identification of violation certificates with
separating portfolios, appears unoccupied.

One principle qualifies everything above.

\textbf{Principle (accessible arbitrage).} \emph{Market coherence is
relative not only to friction but to the cost of discovering violations.
Write \(K\) for the exactly coherent quotes (Proposition 1) and \(K_f\)
for those admitting no sure profit net of a fee \(f\) (Proposition 3).
Let \(\mathcal{A}\) be a class of admissible certificate-search
procedures containing the zero-cost null procedure, let \(x\) encode the
venue's executable books and depth rather than a vector of marginal
quotes, and for a portfolio \(\pi\) let}

\[V_f(\pi; x) \;=\; \inf_\omega\big[\,\mathrm{payoff}_\omega(\pi) -
\mathrm{executionCost}_x(\pi) - \mathrm{fees}_f(\pi)\,\big]\]

\emph{be the guaranteed profit, maximized over the positions the
arbitrageur's depth, capital and position limits allow, with \(C_A(x)\)
the search cost and the expectation taken over the randomness of the
search procedure alone. The accessible-coherence set is}

\[K_{\mathcal{A}, f} \;=\; \Big\{\, x :\ \sup_{A \in \mathcal{A}}
\mathbb{E}\big[\, V_f(A(x); x) - C_A(x) \,\big] \le 0 \,\Big\}.\]

\emph{With nonnegative search costs, \(K \subseteq K_f \subseteq
K_{\mathcal{A}, f}\): a violation persists whenever every available
separation procedure costs at least as much, net of friction, as the
certificate it finds is worth.}

Each feature of \(V_f\) is needed. The supremum ranges over procedures,
not certificates: two procedures can find the same portfolio at very
different costs, so coherence is relative to a class of arbitrageurs.
The null procedure makes the supremum at least zero, so that \(K
\subseteq K_f \subseteq K_{\mathcal{A}, f}\) holds. The profit is a
worst case over outcomes, not realized or belief-weighted profit,
without which exact coherence would not imply that every procedure has
non-positive value.

The profit must also be realizable. A separating functional may be
scaled by any positive constant, so against fixed linear prices its
nominal value is unbounded and no finite search cost could ever leave a
violation standing; it is the nonlinear book, the finite executable
depth, and the position constraint that make
\(K_f \subsetneq K_{\mathcal{A}, f}\) possible at all. Equivalently one
may normalize certificates to \(\lVert y \rVert \le 1\) and price them
against available depth.

Instantiating \(\mathcal{A}\) gives polynomial-time coherence,
bounded-budget coherence, and latency-constrained coherence, and a venue
can be arbitrage-free to ordinary participants while arbitrageable to a
better-equipped class. Three nested notions of coherence result: exact
(no sure-profit chord, Proposition 1), frictional (violations below the
spread survive, Proposition 3), and accessible (violations too expensive
to discover survive). Proposition 11 addresses the existence of
arbitrage; whether a procedure in \(\mathcal{A}\) can profitably prove
\(x \notin K\) is its accessibility, a different property. The
self-policing of the marketization table is qualified accordingly:
effective self-correction is the error's value net of discovery cost and
friction.

Open problems, in rough order of tractability:

\emph{Market representations of general prediction maps.} The
characterization above is confined to cost-based predictors. Define what
it means for an arbitrary prediction map \(T\), from data sets to
forecasts, to possess a market representation, and give conditions on
\(T\) equivalent to existence. A first conjecture: \(T\) is
representable when it factors through a network of cost kernels whose
boundary reduction is \(T\), with the chord bound applying kernel by
kernel; path independence and the chord condition would then be the
image of the general condition in the cost-based case.

\emph{Accessible coherence, characterized.} Compute
\(K_{\mathcal{A}, f}\) for concrete classes: polynomial-time
arbitrageurs against cones whose separation is hard, budgeted
arbitrageurs against fee-bearing venues, latency-constrained
arbitrageurs against cascading settlements. And characterize the
mechanisms in which the stream of certificate payments funds the ongoing
computation.

\emph{The equilibrium fee.} Self-set fees are disciplined by routing,
not set by an equilibrium. In a flow model where most coefficients are
zero and informed trades carry signal plus noise, does the fee that
survives undercutting reproduce the universal threshold
\(\sigma\,(2\log p)^{1/2}\) of the shrinkage literature? If so, the
regularization constant that statistics tunes by cross-validation equals
the adverse-selection cost of the data source.

\emph{Arbitrage depth as a capacity measure.} On a specified compact
domain, with a fixed payoff hull and norm, the arbitrage depth of a
trained model's loss landscape is well defined and plausibly estimable
or boundable, though computing it exactly is a global problem; whether
it tracks quantities learning theory names (sharpness, mode
connectivity) is open.

\emph{From endpoint jumps to quoted depth.} The contact-set analysis
predicts an inventory jump at a single supporting price, and notes the
jump is generally lumpy rather than divisible. The map from that
endpoint correspondence to anything observable in a quoted book is
missing, and is what a test against venues with non-convex liquidity, or
against the bimodal implied densities documented around binary events
(\citeproc{ref-melick1997crude}{Melick and Thomas 1997};
\citeproc{ref-clark2017brexit}{Clark and Amen 2017}), would require
first.

\section*{References}\label{references}
\addcontentsline{toc}{section}{References}

\protect\phantomsection\label{refs}

\end{document}